\documentclass[letterpaper]{article}

\usepackage{natbib,alifeconf}  
\usepackage{url}
\usepackage{comment}
\usepackage{color}
\usepackage[bottom]{footmisc}

\title{On the potential for open-endedness in neural networks\thanks{\hspace{5pt}Final author version as accepted to Artificial Life}}
\author{Nicholas Guttenberg$^{1,2}$, Nathaniel Virgo$^1$, Alexandra Penn$^3$
\mbox{}\\
$^1$ Earth-life Science Institute, Tokyo, Japan
$\quad^2$ Araya Inc, Tokyo, Japan\\
$\quad^3$ CECAN (The Centre for Evaluation of Complexity Across the Nexus) and\\
 CRESS (Centre for Research in Social Simulation), University of Surrey UK}

\begin{document}
\maketitle

\begin{abstract}
Natural evolution gives the impression of leading to an open-ended process of increasing diversity and complexity. If our goal is to produce such open-endedness artificially, this suggests an approach driven by evolutionary metaphor. On the other hand, techniques from machine learning and artificial intelligence are often considered too narrow to provide the sort of exploratory dynamics associated with evolution. In this paper, we hope to bridge that gap by reviewing common barriers to open-endedness in the evolution-inspired approach and how they are dealt with in the evolutionary case --- collapse of diversity, saturation of complexity, and failure to form new kinds of individuality. We then show how these problems map onto similar issues in the machine learning approach, and discuss how the same insights and solutions which alleviated those barriers in evolutionary approaches can be ported over. At the same time, the form these issues take in the machine learning formulation suggests new ways to analyze and resolve barriers to open-endedness. Ultimately, we hope to inspire researchers to be able to interchangeably use evolutionary and gradient-descent-based machine learning methods to approach the design and creation of open-ended systems.
\end{abstract}

The problem of how to achieve open-endedness in artificial systems is a central question of ALife. This is formulated for example as the question of how living systems can generate novel information, as well as how to demonstrate things such as major transitions or the emergence of cognition in artificial systems \citep{bedau2000open}. Despite biological evolution demonstrating the production of a wide diversity of forms across a wide range of scales, modes of interaction, and levels of organization, the artificial systems constructed in mimicry of that process have a frequent tendency to exhibit early saturation --- rather than producing a diverse array of organisms with complex behaviors and forms, they find a small set of organisms which, while they may have some interesting interactions, do not give rise to any further innovations or variations from that point onwards. These tendencies are mirrored in isolated examples from biological evolution, just not the process as a whole, and in those isolated cases it is possible to understand what factors are leading to a limit in the open-endedness of the system, resulting in the development of methods to overcome those limits \citep{dolson2015s}. However, there still exists a fundamental barrier, in that the open-ended systems we produce still do not seem to be capable of indefinitely surprising us, leading to a kind of trivial open-endedness.

In this paper, we would like to look at these limits from the parallel perspectives of evolutionary algorithms and deep neural networks trained via back propagation. Neural networks have a reputation for being less open-ended than evolutionary methods, but we argue that this is mostly due to a difference in the problems which machine learning community has organized around rather than something fundamental to the nature of neural networks or backpropagation. Specifically, the standard of evidence in many machine learning venues is to produce a model which performs better according to a fixed performance metric on a fixed task, which tends to produce a bias against methods which produce a diversity of solutions or which find ways to change the rules of the game. However, recent work with the goal of optimizing for subjective qualities (such as producing photorealistic imagery) has led to exploration of multi-network systems that have more of a co-evolutionary character. These include generative adversarial networks or GANs \citep{goodfellow2014generative} as well as modern reinforcement learning systems such as AlphaGo \citep{silver2016mastering}. We argue that these methods are beginning to achieve some aspects of open-endedness.

There are two crucial issues that are shared between these neural network methods and simulations of open-ended evolution. These are \emph{diversity} and \emph{scaling}. Specifically, once the fixed objective function is removed, it becomes crucial for machine learning algorithms to maintain a diversity of solutions, in order to maintain a memory of solutions or behaviours that were discovered previously. This diversity may occur at the level of outputs from a single network rather than an explicit population, but we argue that the issues involved are nevertheless quite similar. This suggests that ideas about diversity maintenance in genetic algorithms may also be applicable to the training of neural networks, and we outline one way in which this could be achieved. Secondly, if our aim is to increase complexity then we must understand how the pressures to increase or decrease complexity scale with the amount of complexity already present in our system. If this scaling is negative then the complexity will eventually saturate, regardless of how high its initial pressure to increase. We review some recent results suggesting that the scaling laws for large neural networks may be amenable to accumulating information in an open-ended way, with current methods apparently being limited by computing time and the amount of training data, rather than by inherent saturation in the algorithms. This suggests that further study of scaling in neural networks could give new insights about the kinds of systems that are capable of non-trivial open-endedness.

We discuss three issues in particular which interfere with open-endedness both in artificial and real biological systems. In the context of \cite{dolson2015s}, these are ``Novel organisms stop appearing'', ``Organismal complexity stops increasing'', and ``Shifts in individuality are impossible''. The first two of these issues, the tendency of diversity and complexity to both saturate, have been solved in a number of ALife systems, although the solution methods generally constrain the system design. The third issue, the tendency for open-endedness to take forms that are largely 'more of the same', remains a significant problem. While deep neural networks in standard applications tend not to produce a diversity of outcomes, we will talk about how this can be achieved and how the machine learning community is exploring this in the form of the issue of mode collapse \citep{che2016mode, metz2016unrolled, thanh2018catastrophic}. We will also argue that stochastic gradient descent by its nature automatically overcomes the bias towards favoring simpler solutions that generally leads to saturation of complexity. Finally, we will discuss how the tendency of the currently achieved forms of open-endedness to still seem trivial can be linked to the ability of cognitive systems to abstract, and make an argument for how using neural networks as a base may allow us to take advantage of the ability to abstract to make steps towards a more qualitatively satisfying open-endedness.

\section{Diversity}

One sense of open-endedness comes from the observation that biological evolution seems to produce an endless diversity of form \citep{bedau1992artificial}. In this sense, open-endedness just implies that there is always another new form that will be discovered if the system continues to proceed. Yet even this kind of open-endedness can be difficult to produce in artificial systems driven by evolutionary algorithms. The issue is that optimization tends to drive systems to decrease their diversity when in the vicinity of an optimum. Given a set of suboptimal points (genomes or parameters) associated with the same local optimum, those points become compressed together when moved towards the optimum.

In terms of other senses of open-endedness (such as becoming increasingly complex without bound, or never failing to be surprising), a failure to preserve diversity can interfere with the possibility of those other types. A system that can only really maintain one dominant type of behavior or form at a time loses memory of the other behaviors or forms which have been previously discovered. While the system may in the best case manage to move between these attractors, it cannot maintain a long-term preference for new attractors over old ones, and so is not driven to explore. 

In evolutionary systems, the tendency for diversity to collapse takes the form of competitive exclusion \citep{gause1934struggle, hardin1960competitive, capitan2015similar}, in which population dynamics in the relative exponential growth regime will drive all but the highest fitness species to (relative) extinction. Even in the presence of mutation, systems with an Eigen error threshold \citep{biebricher2005error} have a phase transition between a regime in which selection wins out over mutation and the entire population is clustered around a local optimum in the genetic space, and a regime in which mutation ends up erasing the evolutionary history completely --- essentially, preventing the population from successively accumulating information about the environment via selection. Additionally, even in co-evolutionary cases in which the evolutionarily stable strategy can correspond to a heterogeneous population (a mixed strategy), other considerations such as finite population size effects or the distributional details of how populations of one generation map to the next may limit the actual sustainable diversity of the system \citep{ficici2000effects}. Detailed aspects of the dynamics can lead to a failure to achieve even theoretically optimal and stable open-ended evolutionary outcomes \citep{lindgren1992evolutionary}. 

In the next two subsections, we outline how these notions of diversity can map to the training of machine learning models. We first introduce the notion of a generative modelling task, in which a network must maintain a diversity of possible outputs rather than learning a single output for a given input, and then we show how a version of Lehman and Stanley's (\citeyear{lehman2010revising}) minimal criterion novelty search algorithm can be applied to train a network to perform such a task. The purpose of this is to show how concepts can be mapped between the two fields, rather than to present major results. In the remainder of the section, we discuss how these ideas apply to Generative Adversarial Networks (GANs), in which two co-trained networks behave in many ways like two co-evolving populations.

\subsection{Generative modeling tasks}

In machine learning, given that often the focus is on producing a single trained model rather than a population of models, there are various ways one could think of the analogue of 'diversity'. One possibility is to think of the set of possible outputs of a model given a particular fixed input as representative of a population\citep{moran2018coevolutionary}. In this case, classic instances of supervised learning --- that is to say, optimizing the average of a pointwise scalar objective function over the dataset --- necessarily suffer from diversity collapse at their global optimum. Specifically, we can consider a model trained to minimize some function $L = \sum_i f(y_i,\hat{y}_i(x_i))$ where $y_i$ is a target value, $\hat{y}_i(x_i)$ is the output of the model given some associated input $x_i$, and the sum is over the dataset. We can rewrite this:

\begin{equation}
L = \int \mathrm{d}x \mathrm{d}y \mathrm{d}\hat{y} \rho(x,y) \rho(\hat{y}|x) f(y,\hat{y})
\end{equation}

where the $\rho$ functions are the empirical distributions associated with the dataset ($\rho(x,y)$) as well as the corresponding possible outputs the model $\rho(\hat{y}|x)$. If we consider each case of $x$ on it's own, we can rewrite $\rho(x,y) = \rho(x|y)\rho(y)$ and place the $y-$dependent terms into an inner integral. This allows us to obtain a function

\begin{equation}
\tilde{f}(\hat{y}) = \int \mathrm{d}y \rho(y) f(y,\hat{y})
\end{equation}

such that:

\begin{equation}
L = \int \mathrm{d}x \mathrm{d}\hat{y} \rho(x) \rho(\hat{y}|x) \tilde{f}(\hat{y})
\end{equation}

In essence this says that, given the distribution of possible true values $y$ for a particular $x$, we can replace the distribution of contributions to the overall objective function made by those $y$ values with a single characteristic value that averages over the dataset. Assuming the model is sufficiently flexible to do so, the optimal solution for this which minimizes $L$ is always for $\rho(\hat{y}|x)$ to be a $\delta$-function around some particular $y_*(x)$ associated with each unique input.

While this may be true for the usual type of supervised learning task, there is a family of tasks where the goal is not to output a specific thing in response to a particular input, but rather to be able to learn to generate samples from a particular associated distribution. These are refered to as generative modeling tasks, and in such cases the objectives are constructed in different ways so as to be able to take into account the distribution of outputs (and as a result, to be able to converge to an optimal and yet diverse set of output behaviors). This can be done by learning a transform from a starting distribution which is easy to sample from (such as a high-dimensional Gaussian) into the target distribution, by learning to estimate the likelihood of a given point, or by explicitly outputting summary statistics of a distribution model (e.g. in the case of a Gaussian mixture model, this corresponds to the means, covariance matrices, and relative weights). It is also possible to make autoregressive generative models that iteratively sample one dimension at a time from the target space, conditioned on the dimensions which have been generated so far. In each of these cases, the actual stated objective function cannot be generally satisfied by a non-diverse output.

\begin{figure}
 \includegraphics[width=\columnwidth]{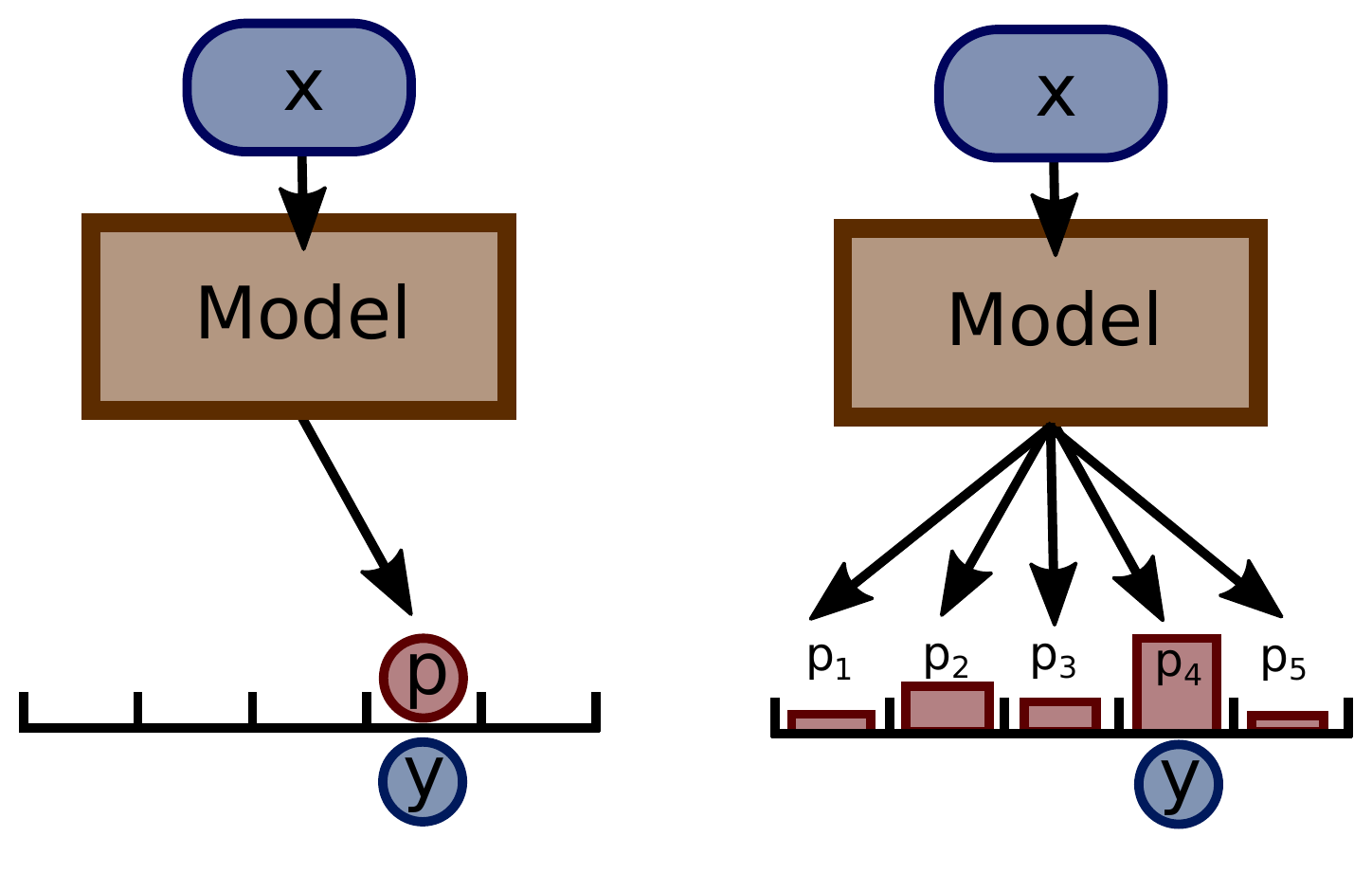}
 \caption{\label{PClassifier} Left: a model which generates a single (sampled) prediction given a particular context. Right: a model which parameterizes a probability distribution over possible predictions given a particular context. }
\end{figure}

The simplest example of such constructions is when a model is trained to explicitly output the components of a probability distribution (Fig.~\ref{PClassifier}). In multi-class classification, this is done in practice by placing a softmax activation function after the final output of the rest of the model. The softmax function maps its input vector $\vec{x}$ to a vector $\vec{p}$ of positive values $p_i$ which are guaranteed to sum to unity, thereby allowing the arbitrary vector $\vec{x}$ of input values $x_i$ to be interpreted as parameterizing a probability distribution $\vec{p}$. Formally the components of $\vec{p}$ are defined via:

\begin{equation}
p_i:= \textrm{Softmax}(\vec{x})_i = \frac{ \exp(x_i) }{\sum_j \exp(x_j)}
\end{equation}

Then, the model is trained to minimize the categorical cross-entropy (CCE) between the output probability distribution $\vec{p}$ and the true probability distribution $\vec{y}$.

\begin{equation}
 \textrm{CCE}(\vec{p},\vec{y}) = -E[ \sum_i y_i \log p_i  ]
\end{equation}

where here the sum is over the different possible class labels. The CCE is a function which measures the divergence between two distributions (here $\vec{p}$ and $\vec{y}$), and is minimum when those distributions are equivalent. Since the true probability distribution is generally only estimated from data, in effect for a given set of observations this becomes:

\begin{equation}
 L = - \frac{1}{N} \sum_j^N \log p(y_j | x_j)
\end{equation}

where now the sum is over individual instances $y_j, x_j$ from the dataset. While this is still a point-wise comparison between samples, the output does not consist of samples but rather describes the components of a probability distribution. As a result, while the model would optimally converge to a single particular output given a particular input, that output can still represent something with non-zero entropy and the corresponding distribution $\vec{p}$ can be sampled from.

This approach can be broadened by thinking of the components of $\vec{p}$ as simply a way to parameterize some probability distribution with a known analytic form (in this case, a piece-wise constant function). Other distributions can be used instead, such as in the case of Mixture Density Networks \citep{bishop1994mixture} which have the model output the parameters for a set of multiple Gaussian distributions.

The general use-case for this sort of approach is when there's a reasonable guess for the family of distributions that will model the data well, or when the outputs are in a sufficiently low-dimensional space that the distribution can be discretized over a grid.

\subsection{Minimum-over-set losses}

Much like the minimal criterion method for genetic algorithms \citep{lehman2010revising}, it is possible to obtain a diversity of outcomes from a neural network using losses which only ask for the network to achieve some sufficient result rather than an optimal result. This technique is used in Time Agnostic Prediction \citep{jayaraman2018time} to make a network which attempts to predict future frames from a video sequence, but which essentially has a choice as to which frame it will try to predict. This is done by allowing the network to not be penalized for bad predictions in frames other than the one where the prediction is best --- that is to say, it is sufficient for the network to predict one frame well. 

This type of sufficient criterion can be extended to the creation of a basic generative model where the network converges to a diverse distribution of outputs. The basic structure of the idea is to take a network which operates on some input $N(x)$ and augment it by adding a input in the form of a sample from a noise distribution $\eta$: $N(x,\eta)$. If the network is being trained against some loss function $L(N(x),y)$, then rather than minimizing the loss function directly, one can instead train against:

\begin{equation}
 \tilde{L} = \min_\eta L( N(x,\eta), y )
\end{equation}

\begin{figure}
\includegraphics[width=\columnwidth]{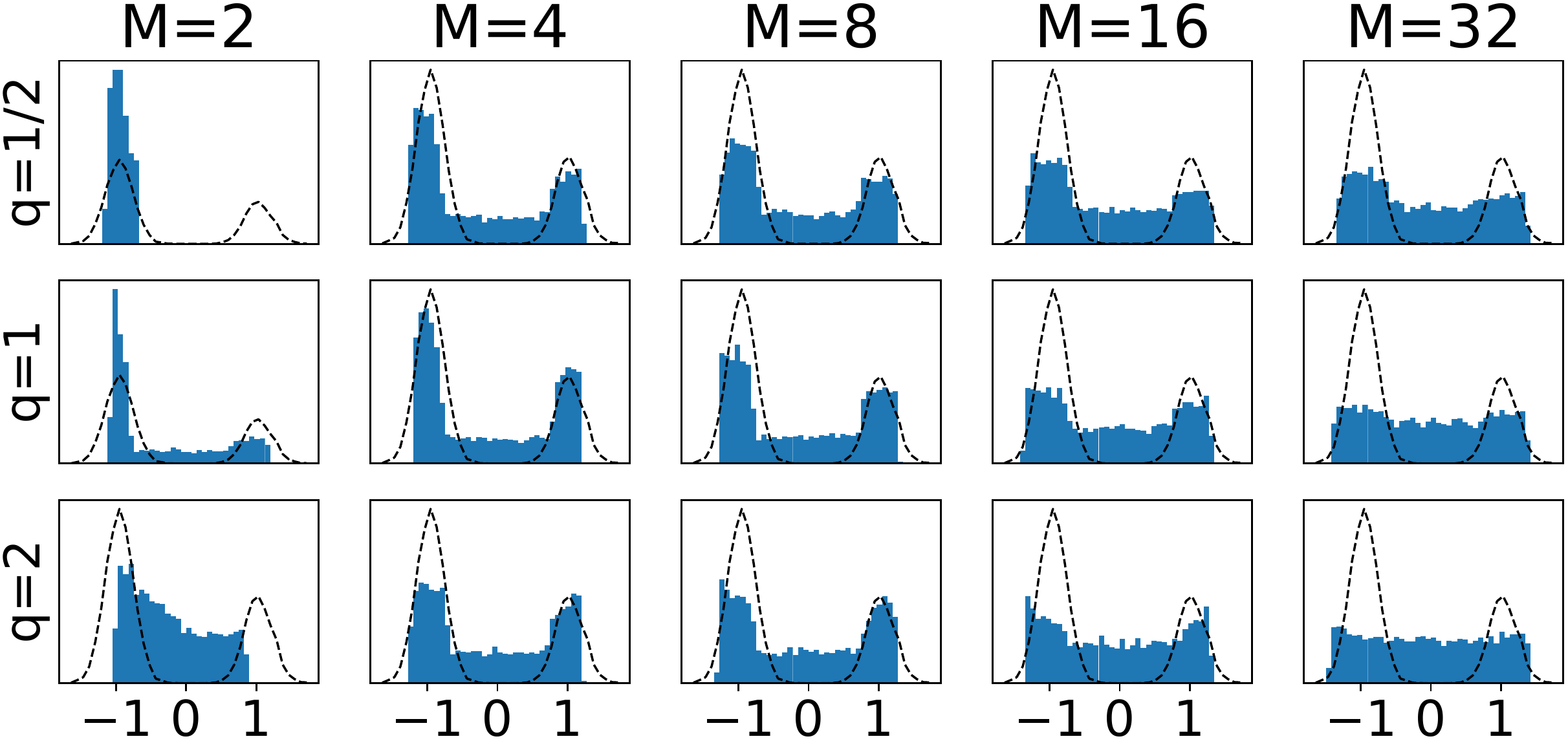}
\caption{\label{Minloss}Generated distributions of networks trained with a minimum-of-$M$ loss, of the form $L(x,y)=|x-y|^q$. For large $M$, $q$, the network converges not to the data distribution (dashed line) but rather to its support.}
\end{figure}

That is to say, the network is asked for the target value $y$ to be within the support of the distribution $N(x,\eta)$, rather than just being asked to output the target value directly given $x$. For any values of $\eta$ which do not correspond to this best-case value, there is no direct optimization pressure even if the values are very far from the target value. In practice, since one would only use a finite number of samples from $\eta$, the network is encouraged not to 'waste' values of $\eta$ on values of $y$ which never show up. This has the consequence that, much like with the categorical cross-entropy, if there is some uncertainty in $y$, the optimal solution is to output a distribution of values that cover what $y$ could conceivably be for any particular $x$, rather than to converge to a single point output.

Depending on the precise loss used for $L$, the relationship between $p(N(x))$ and $p(y)$ can be more complicated than just equality (Fig.~\ref{Minloss}). If for example $L(x,y) = |x-y|^q$ , then for large values $q$ (and correspondingly, large numbers of samples $M$ from $\eta$), $p(N(x))$ is driven to become constant in areas where $p(y)$ is greater than some threshold and zero elsewhere --- when $q$ is large then samples only need to be within a certain radius of the target value, and when $M$ is large then low probability events in $p(N(x, \eta))$ are mapped to higher effective probabilities in the expectation: $\tilde{p}(N) = 1-(1-p(N(x,\eta)))^M$ for the best value of $N$. On the other hand, as $q \rightarrow 0$, and as $M$ becomes small (but still $>1$), the model must more densely cover the peaks of the distribution at the cost of the tails, because near-misses count for less and less.

This sort of approach does not scale well to very high-dimensional output spaces, since the probability of finding a good value of $\eta$ which causes the network to land close to the target $y$ gets smaller as the dimension of the output space increases. So we present it here mostly as an observation that the same intuitions which can be used to drive diversity in genetic algorithms can also apply to driving diversity in the outputs of neural networks.

\subsection{Generative adversarial networks}

The previous two techniques are used to train single networks which can stably converge to outputting distributions of outcomes. However, the dynamics of training is fundamentally still convergent towards an optimum. If divergent behavior is necessary for deeply exploratory open-endedness (something proposed as a role of mechanisms such as mutation), then the above techniques would not suffice.

\begin{figure}
 \includegraphics[width=\columnwidth]{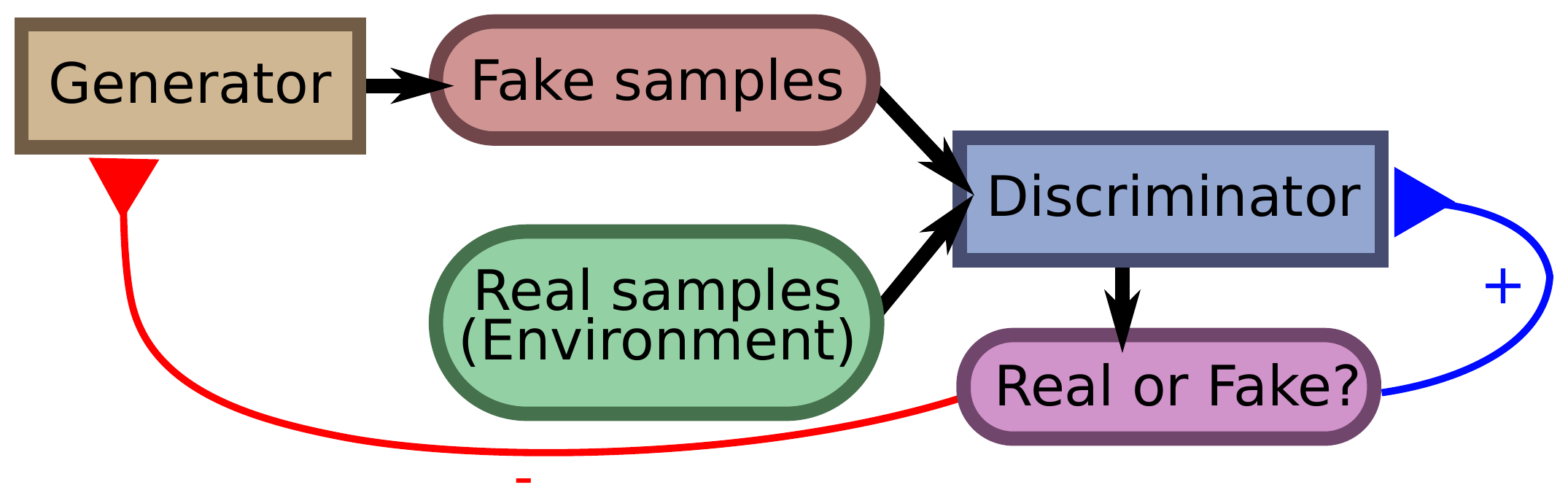}
 \caption{\label{GAN} Sketch of the GAN architecture. The generator produces 'fake' samples and is rewarded for fooling the discriminator. The discriminator classifies samples as real or fake, and is rewarded for doing so correctly. }
\end{figure}

On the other hand, the method of Generative Adversarial Networks (GANs) (Fig.~\ref{GAN}) implements a coevolutionary arms race, which (in the biological equivalent at least) does produce divergent behavior. GANs consist of two networks: a generator and a discriminator. The discriminator is trained to classify samples as either belonging to the data distribution or being from the generator, while the generator attempts to produce samples which can fool the discriminator. Overall, the method is motivated by the observation that the Nash equilibrium of the competitive dynamics should be that the distribution of the generator's output is exactly matched to the data distribution. However, since the two networks are trained with opposing loss functions, there is no strong guarantee that the overall dynamics will converge directly to that equilibrium. In fact, a recent systematic study across GAN variations shows that even when the generated samples appear to have converged, in reality the network weights may be exhibiting behavior such as limit cycles in the vicinity of the Nash equilibrium \citep{mescheder2018training}. This suggests an analogy to evolutionary game theory, in which the instability of Nash equilibria and possibility of limit cycles are also important considerations.

Conceptually, GANs get closer to the idea of open-endedness than the other methods in the sense that they can operate in a much higher dimensional space than can be exhaustively mapped. Autoregressive models try to capture those high dimensional structures by explicitly factorizing the space into independent distributions, while using a minimum loss over multiple samples requires the samples to span the space meaningfully in order to capture details. In a GAN, the discriminator network effectively searches for an interesting direction --- defined in the sense that the generator has yet to correctly capture something about the data in that direction. Then, following that, the generator exploits the discovered direction and fixes that aspect of the distribution. Since these 'directions' are constructed from deep representations in the discriminator, they need not correspond directly to microscopic degrees of freedom or details, but can instead capture higher level abstractions about the data and its internal relationships. As such, GANs can capture things such as a sense of photorealism as meaning 'perceptually indistinguishable from reality' rather than literally having the same pixel values as a particular photo.

When in a stable training parameter range this iterative procedure of finding some inconsistency and fixing it ends up capturing all such inconsistencies which can be detected. As such, even though the objective function is relatively simple, the networks can in principle capture as much diversity as exists within their environment (e.g. the training data). In practice, however, GANs exhibit a phenomenon known as 'mode collapse', in which the generator cannot stably maintain coverage over the entire distribution, but rather jumps from place to place, modelling a succession of sub-parts of the data. 

One explanation for mode collapse is that, while the Nash equilibrium between the networks should contain the full entropy of the data distribution, if one were to hold the discriminator fixed and trained the generator to completion, for any simple scalar loss function this should always result in only a single best sample being generated \citep{goodfellow2014generative, metz2016unrolled}. In the actual joint training, the generator never collapses completely to a single sample due to the movement of the discriminator, but the general pressure is to converge on that point. However, if rather than training the generator to fool the current discriminator, the generator were trained to fool a converged future discriminator (that is to say, if the generator were trained to make it hard for the discriminator to win), then the optimal solutions are not pure samples but are instead distributions. This method, referred to as unrolled GANs \citep{metz2016unrolled}, is much more stable against mode collapse.

Looking at this strategy, there is a commonality with the minimum-over-set losses. When using a set of samples and optimizing only against the best case from the set, the optimum solutions become distributions rather than pure samples because it becomes advantageous to generate bad samples that have some chance of being good samples under some circumstance. With the unrolled GAN, rather than taking the best sample for a fixed context out of a random set of samples and optimizing it further, one is essentially generating a fixed set of samples but then finding the worst possible single context (by considering a future adapted discriminator) across that set. As a result, not only should the generator hedge its bets (which is effectively true anyhow with regards to stability at least), but the loss function that the generator optimizes directly takes this into account.

In terms of evolutionary dynamics, this suggests a potential interaction between minimal criterion based fitnesses and the Baldwin effect \citep{baldwin1896new, fernando2018meta}. Much like these neural networks, organisms with methods of within-lifetime adaptation can express a diverse set of phenotypes even if genetically identical. While a minimal criterion (such as different organisms having different bottleneck resources in a system with multiple niches) supports a distribution of genomes, even when evolutionary pressures such as competitive exclusion would prevent genetic diversity from being stable, within-lifetime mechanisms of adaptation can take over and provide that diversity directly in the phenotype.

\subsection{Bounded diversity of generative models}

Ultimately, generative models as used in the machine learning community still have effectively bounded diversity as their objective is to match some particular fixed distribution of data (although, if that data comes from the natural world, the diversity may be correspondingly high). This is a place where the specific goal of producing a numerical model which is capable of generating or predicting something concrete about the world is likely to be creating a bit of a blind spot. An exception to this is perhaps in the field of reinforcement learning, where it is often necessary to make agents to explore the space of the possible in a given environment.

The method of density estimation-based curiosity \citep{ostrovski2017count} implements a kind of exploration algorithm based on an underlying generative model. Specifically, an autoregressive generative model is learned for the states that an agent visits, and then behaviors which lead to low-density regions of the probability distribution are reinforced. Much like novelty search \citep{stanley2011}, this leads to agent behaviors that try to extract a maximum variety of outcomes from the environment. Of course, now rather than the diversity of the method being bounded by the data, it is bounded by the diversity of sensor values that the agent's particular environment can give rise to, which is generally strongly bounded in the sorts of games that are used to test this kind of method as the game itself is not something which has any way of varying.

Multi-player games on the other hand can give rise to emergent complexities and variations due to the need to not just adapt to the rules, but also to adapt to the strategies of the other player. We will discuss this more in the following section, specifically with regards to the idea of self-play between neural networks and copies of themselves, a technique that exploits the sustained pressure produced by competition to explore new strategies in order to accelerate learning.

The intersection between these ideas may provide a fertile ground for expanding the scope of meaningful, open-ended diversity generation. Rather than training a generative model against the world, generative models could be connected together and trained to produce transforms of each-others' outputs related by what amount to rules for interaction (or rules of a game which they share). The adaptation of GANs to modelling populations of agents playing Prisoner's Dilemma and other basic games \citep{moran2018coevolutionary} is a step in this direction.

\section{Complexity and scaling}

Indefinite diversity production covers one idea of open-endedness, but there are many examples of trivial processes which produce output distributions of arbitrarily high entropy. For example, neutral point mutations of an infinitely long non-coding section of a genome might end up covering a space of infinite potential diversity, but such diversity ultimately has no relationship with the organism's behaviors or with the context in which it exists --- it is non-functional. It is also possible to imagine spaces which are effectively infinite in their capacity to express functional diversity, but where the choices as to what particular thing to express are essentially arbitrary. An example of this is the naming game \citep{steels1998origins} in linguistics, where for $N$ concepts and potential words, there are $N!$ possible languages which map a single word to a single concept and as such would be functionally equivalent to each other.

This leads to the idea that not only should an open-ended system produce a diversity of possible outcomes, but those outcomes should be progressing in some direction such that the newest outcomes depend on and extend what was produced in the past, in analogy to things such as the development of technology. This can be compared to our argument in the previous section, where we argued that diversity can be important as a way of providing a memory of previously discovered solutions or behaviours.

This idea of directionality is often summarized as the idea that things produced by the system should grow increasingly complex over time, where the notion of complexity could have a variety of different concrete interpretations --- Kolmogorov complexity \citep{kolmogorov1965three}, information content with respect to the world or other agents, interdependency of components with respect to function, etc. Very broadly, a common element to these measures is that the particular details of things should matter: if one shrunk an organism's genome to the size of the Kolmogorov complexity of its behaviors, then not one change could be made without destroying the behaviors; if one did the same with respect to the mutual information instead, then every change made is a loss of some potential for the organism's behaviors to relate to its environment; if one of a set of interdependent components is altered, it influences the function of all the others; etc.

It is easy to find systems in which such quantities are driven by selection pressures to increase, but this is separate from the question of what is necessary for them to increase without bound. Effects which are insignificant at finite scale may become dominant when some aspect of the system is diverging towards infinity. As such, it may often be necessary for certain effects or considerations to have exactly zero effect, rather than just a sufficiently small effect, in order to preserve the divergence. Resonance in oscillators is an example of this, where even an infinitessimally small amount of dissipation in the system becomes the dominant effect determining the shape of the resonance as well as the peak amplitude that will be observed under a given energy input. Similarly, in the Ising model critical phase transition, any small non-zero external magnetic field is sufficient to detune the system from criticality, and as a result becomes a dominant effect in controlling the cutoff of the divergence of the specific heat or magnetic susceptibility around the transition.

\begin{figure}
 \includegraphics[width=\columnwidth]{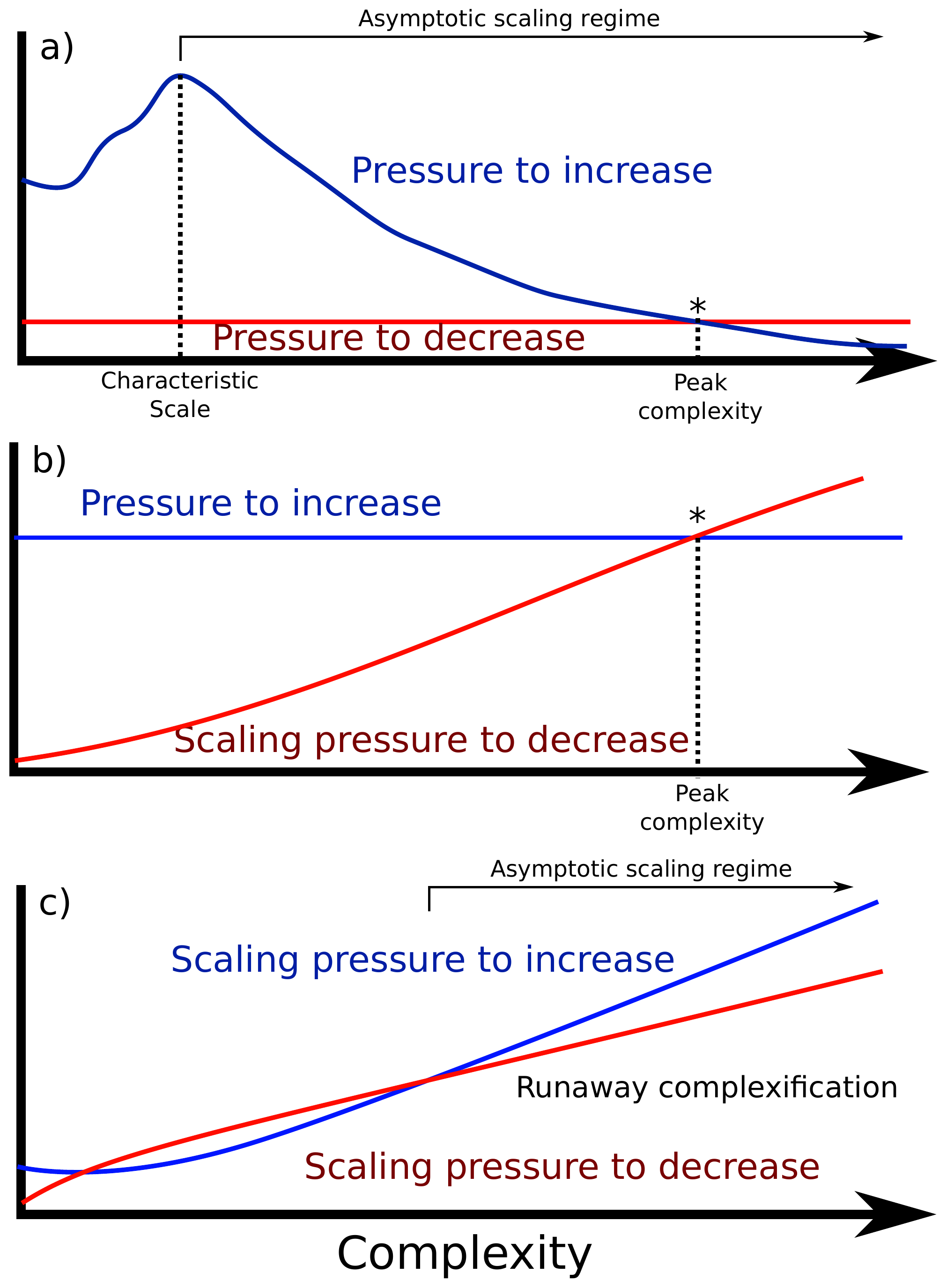}
 \caption{\label{CScaling} a) Complexity peaks because of diminishing returns on pressure to increases, versus static pressure resisting. b) Complexity peaks because of scaling pressure to decrease versus consistent static pressure to increase. c) Pressure to increase complexity grows asymptotically faster than pressure to decrease, leading to runaway increase in complexity.
 }
\end{figure}

In terms of an open-ended increase in complexity, the sorts of terms which might detune a divergence are scaling costs or diminishing returns such that, as the complexity of the system increases, either the cost of maintaining that complexity becomes divergently large (compared to other forces on the system), or the forces pushing the complexity to increase become divergently small (compared to fixed forces preferring a particular complexity scale). Such forces might include selection pressures, but they can also include mutation biases or the entropic cost of maintaining long sequences, for example. An example of this would be that when there is some fixed external task which in part determines an organism's fitness, then there is also likely a particular associated characteristic complexity scale beyond which the strength of selection pressures associated with improving that fitness via increasing the complexity will either decay due to diminishing returns or even reverse. If there is even a fixed-strength, extremely weak pressure to decrease complexity present in the system, that pressure will win in the infinite limit against a decaying pressure towards increasing complexity (Fig.~\ref{CScaling}a).

Instead, in order to preserve pressure towards complexity increases, it may be necessary for any selective benefits to be relative to the complexity scale of the current population, rather than corresponding to an absolute, fixed landscape. Coevolutionary dynamics are one way of providing this sort of effect. In essence, a coevolutionary system produces a (potentially un-ending) sequence of successive tasks. Depending on the structure of the interactions, this may indefinitely provide a consistent local selective benefit for increasing in complexity relative to the other organisms in the system. An example of this is runaway selection effects in red queen dynamics. For instance, plants competing for sunlight would receive a selective advantage only if they manage to escape eachothers' shade, regardless of the particular height at which that is achieved. So by making the dominant pressures relative rather than absolute, those pressures can persist even as the complexity of evolved structures diverges.

Even if the pressure towards increasing complexity does not decay, a growing pressure or cost associated with maintaining an increasing repertoire of information could also lead to saturation of the complexity that a given system achieves (Fig.~\ref{CScaling}b). Such a cost could arise from the need to implement repair mechanisms to reduce the effective mutation rate, or as an overall reduction in fitness associated with an increase in the number of potential deleterious mutations which could occur in an organism's offspring. Phenomena such as Spiegelman's Monster \citep{spiegelman1965synthesis} and 'survival of the flattest' \citep{wilke2001evolution} reflect these considerations in the form of an intrinsic evolutionary bias towards reducing genetic complexity as much as possible while satisfying the constraint of being able to survive and replicate. These phenomena have also been observed in artificial systems of replicating programs such as Tierra \citep{ray1992evolution}, in which the addition of normalization by execution time was needed to avoid an implicit bias towards shorter replicators. In general, it seems that complexity fundamentally comes at some cost, be it in fitness or in robustness. In order for complexity to increase without bound, there must be ways in which that cost is kept in check as complexity increases, in which the net benefit of increasing complexity is kept above the level of the increasing cost, or in which reductions in complexity are made non-viable (complexity ratchet) \citep{complexityratchet}.

The degree to which systems can obtain an arbitrarily high complexity is then bounded by the degree to which such mechanisms can scale without reaching a hard cutoff. Eigen showed that under a particular rate of genetic drift, there is a maximum amount of information which can be maintained in a given genome at a particular mutation rate \citep{Wallace2009}. This results in the so-called Eigen Paradox, where the amount of information needed to code for genetic repair mechanisms which effectively lower the mutation rate is greater than can be sustained without already having those mechanisms in place. In order for the amount of information in an organism to diverge to infinity (that is to say, to actually be able to continue to increase in complexity without ever stopping), the effective rate of point mutations must decrease to zero. In the context of characteristic scales suppressing criticality, we can understand this by observing that point mutation has a characteristic scale in the form of the unit of information storage which is being mutated. When the sorts of structures which encode the function of an organism become asymptotically larger than the point mutation scale then there is a commensurate diverging entropy cost for maintaining those structures, which acts opposite to selection pressure.

If on the other hand the mechanisms of evolutionary drift are also scale-invariant or are coupled to the organism's complexity, then this can avoid the existence of a cutoff scale. For example, if increases in complexity allow proportionately better repair of mutations, such that the effective mutation rate per offspring that can be achieved decreases at least inverse linearly with the amount of genetic information stored, then the system can at least in principle indefinitely stay ahead of the error threshold. Other forms of genetic variation such as horizontal gene transfer, lack a characteristic scale and therefore can sustain indefinite increases in complexity. 

\begin{figure}
 \includegraphics[width=\columnwidth]{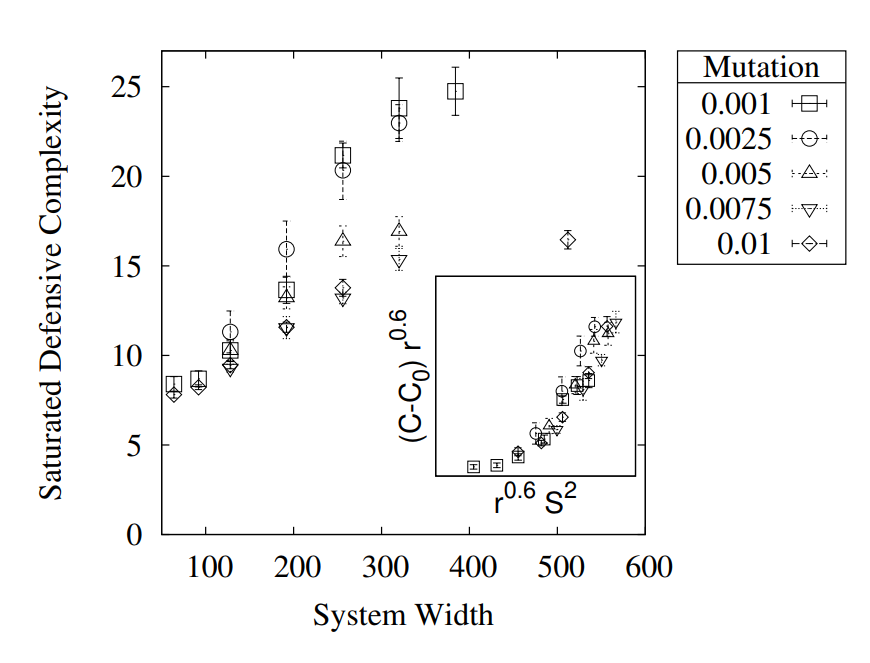}
 \caption{\label{Datacollapse} Reproduction of Fig.~3 of \cite{guttenberg2008cascade}. This shows the scaling of organism complexity with respect to system size (S) and point mutation rate (r) for a predator-prey system. The inset depicts evidence for the existence of a critical point associated with this data, in the form of an observed data collapse when the data is plotted against combined power-law functions of the mutation rate, system size, and complexity.}
\end{figure}

In evolutionary systems, we previously showed that it was possible to drive various complexity metrics to increase indefinitely by way of a general recipe of suppressing the existence of characteristic scales in the evolutionary dynamics, and then applying relative selection pressures. This was applied to three systems. In one, we used competitive predator-prey dynamics with an attack-and-defense motif, where attackers would need to find some pattern not covered by the defender in order to successfully eat them \citep{guttenberg2008cascade}. This caused the fitness landscape to be entirely constructed out of comparison to the rest of the ecology, with no 'fixed' terms associated with particular complexity scales. Furthermore, point mutations were augmented (and asymptotically replaced) by a scale-free gene duplication dynamic. This recipe was extended to a symbiotic version of the same system, where organisms would elect to consume compounds from their neighborhood and would emit byproducts which would require a slightly more complex process to subsequently metabolize \citep{guttenberg2009scale}. While the competitive system produced dynamics much like a travelling wave in sequence complexity (with both attackers and defenders peaked near the maximum complexity achieved by the ecosystem so far), the symbiotic system produced a diverse distribution of organisms spanning an increasing range of trophic levels. We also investigated a system of plants learning to encode 3d morphologies in order to compete for sunlight, and found similar results as the competitive predator-prey system \citep{guttenberg2009scale}. In particular for the predator-prey system, we observed that the increase in complexity would saturate at levels dependent on the residual point mutation rate and system size, in a fashion consistent with the existence of a critical point at infinite size, zero point mutation rate, and infinite organism coding length (Fig.~\ref{Datacollapse}). 

The constraint of suppressing the existence of characteristic scales is a strict one, as it limits things to systems based on rules which are sufficiently self-similar that the scaling behavior of the system can be guaranteed. It is not clear that, for example, the space of arbitrary programs should have such convenient self-similarity properties. As such, when something like Tierra demonstrates an asymptotically saturated complexity, it is difficult to know whether that is because of the scales introduced by mutation operators, by the fitness function, by some implicit property of the environmental dynamics, or due to some structural properties of programs represented in that language.

This makes neural networks an interesting space to work in with respect to open-endedness of complexity. Many classes of neural networks have been shown to be universal function approximators \citep{cybenko1989approximations, hornik1991approximation}, in the sense that there is always some finite sized neural network that can approximate a given function to an arbitrarily small error rate on a particular set of data. Furthermore, it has been observed that neural networks are able to perfectly fit structureless noise patterns given sufficient training time --- meaning that not only are arbitrary functions contained within the domain of their expressiveness, but that the training process itself can actually successfully find such functions. This gives us something like a space of programs where apparently the encoding of such programs does have the right sort of invariances so that arbitrarily complex things are not a priori excluded from discovery.

\subsection{Scaling in Neural Networks}

\begin{figure}
\includegraphics[width=\columnwidth]{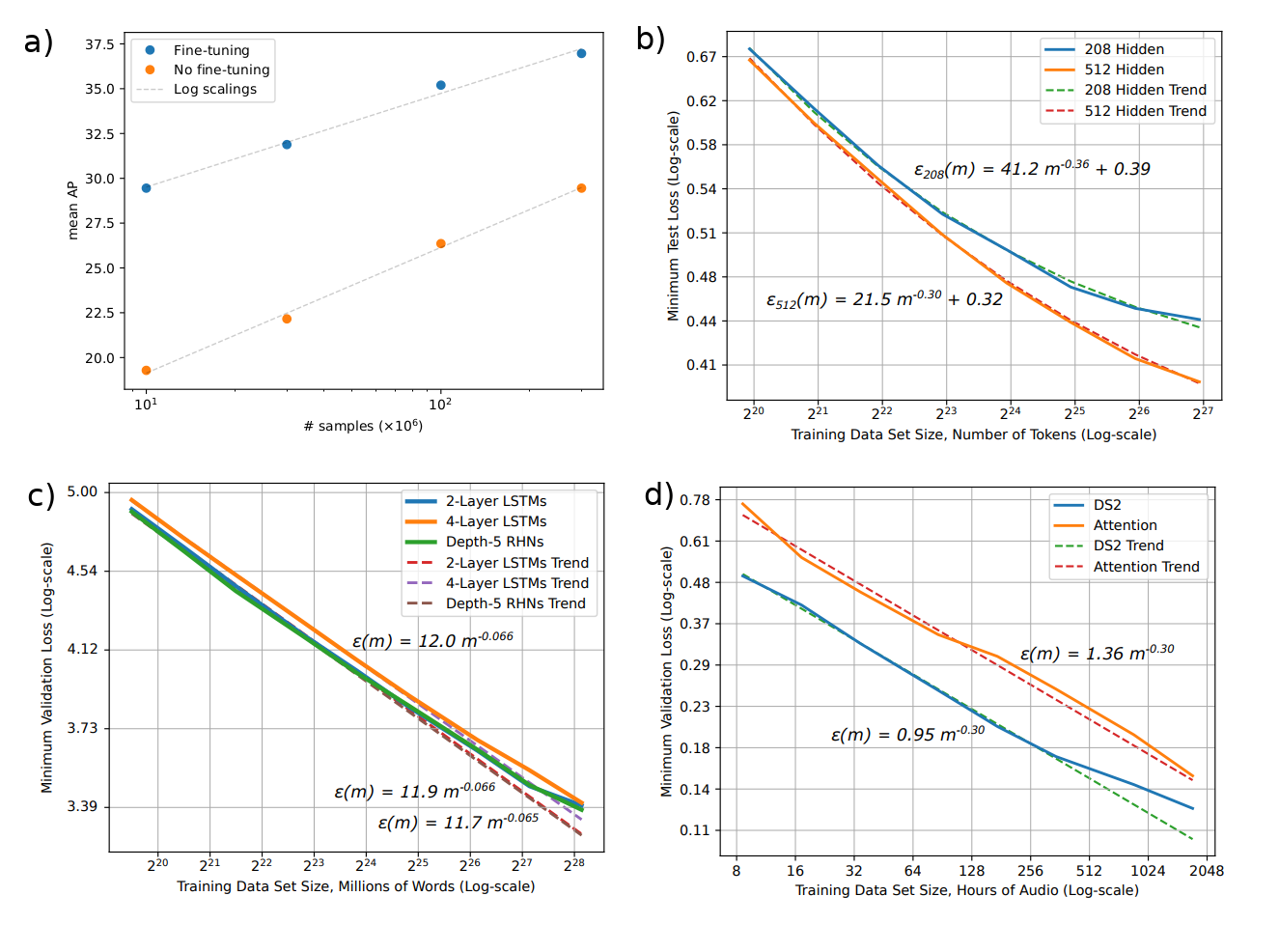}
\caption{\label{Scaling} a) A replot of the data from Fig. 4 (left) of \citep{sun2017revisiting}, showing logarithmic scaling of network performance with number of training examples. b) Fig. 1 (left) from \citep{hestness2017deep} showing power-law scaling of error with data in a translation task. c) Fig. 2 (left) from \citep{hestness2017deep} showing power-law scaling of error with data in a language modelling task. d) Fig. 5 (left) from \citep{hestness2017deep} showing power-law scaling of error with data in a speech recognition task.}
\end{figure}

First we will examine a body of empirical evidence as to the scaling properties of neural networks that has emerged from commercial applications over the last few years. Neural networks have now been trained on image classification, speech recognition, and natural language modelling and translation across many orders of magnitude of available data and network size, and appear to demonstrate consistent scaling of performance over that range. The primary limiting factor is that in order for performance to continue to grow with additional data, the network size must also be increased, strongly suggesting that this is driven by the network's ability to include more information rather than just the optimization of model parameters becoming more precise. 

In the image domain, scaling experiments on a dataset consisting of $3\times10^8$ training images and 18291 categories (with multiple categories per image) \citep{sun2017revisiting} showed consistent scaling laws of network performance with respect to data. The mean average precision, which measures the overlap between the predicted categories and the actual category list, was observed to increase logarithmically as the amount of training data was increased over range between $10^7$ and $3 \times 10^8$ examples, so long as the network size was sufficiently large (Fig.~\ref{Scaling}a). More recently, a study \citep{mahajan2018exploring} on the effects of network pretraining made use of a $3.5\times 10^9$ image Instagram dataset with noisy labels, in order to pretrain a network before fine-tuning and testing on the standard ImageNet benchmark \citep{deng2009imagenet}. In this case, they observed that pretraining a sufficiently large network produced consistent logarithmic improvements in performance with respect to data quantity over the range from $10^7$ to $3\times 10^9$ images, so long as the target task was sufficiently difficult.

Since accuracy is a bounded measure, it may be more appropriate to look at the decrease of error rather than the increase of accuracy. Power-law scalings with respect to data size have been observed from $3\times 10^4$ to $5\times 10^5$ images on ImageNet data, $8$ to $2048$ hours in speech data, and a range of roughly two orders of magnitude in various language modelling and translation tasks \citep{hestness2017deep} --- corresponding figures reproduced in (Fig.~\ref{Scaling}b,c,d). These power-law scalings come with commensurate power-law increases in network size necessary to avoid saturation with respect to the amount of data, and range from $~N^{-0.3}$ scalings in the image and speech recognition domains to a $N^{-0.066}$ scaling in word-level language modeling.

These results seem to show that as long as the external context of the network (that is to say, the tasks and data on which it is trained) has additional structure to be gleaned and there is sufficient data available, even when there are diminishing returns, asymptotically infinite neural networks trained using stochastic gradient descent and backpropagation are able to learn and incorporate that information successfully. So empirically at least, it appears that the underlying learning mechanisms for training neural networks do not suffer from a characteristic information retention scale which would force saturation if they were otherwise driven towards infinite complexity. At the same time, these are all cases in which the driver for complexity corresponds to an external process which must presumably already possess as much complexity as the corresponding network which would be trained to model it.

\subsection{Complexity of Neural Networks}

To better understand these scaling results, we can consider a formal sense of 'complexity' with respect to the capacity of statistical learning to differentiate between different models, used to compare a wide variety of machine learning methods. The basic idea is that if a model is sufficiently expressive to fit arbitrary data, then the fact that the model succeeded or failed to fit a particular dataset provides no information as to whether that model describes the true underlying process which generated the data, versus just memorizing the particular data samples as given. One such measure of model expressivity is the Vapnik-Chervonenkis dimension (VC dimension) \citep{vapnik2015uniform}, which is the number of data points that are guaranteed to be able to be 'shattered' by choosing optimal values of the model parameters. 

Within traditional machine learning techniques, the VC dimension is used as follows: Given a constrained model family which contains a 'true' model that obtains a minimum error on the underlying process which generates the data, the VC dimension puts a bound on how badly a given model will generalize to new samples. This classical result essentially says that if a model is better at fitting the data, it will always be worse at generalization. However, empirically it seems this is not necessarily true for large neural networks. In this section we briefly review this paradox, which is not yet well understood and may provide insights into the kinds of system that can learn in an open-ended manner.

The asymptotic scaling is that essentially, a number of observations linear in the VC dimension of a model family are needed in order to maintain a constant generalization bound. So whereas the case of point mutations required us to reduce the mutation rate asymptotically to zero to obtain arbitrary complexity, there also appears to be a limit where the number of observations needed to establish a 'meaningful' complexity (e.g. one which is not just a product of happenstance) must diverge to infinity as the desired complexity diverges to infinity (which, in terms of this sort of bound, should conceivably apply both for evolutionary processes and machine learning processes, as it does not distinguish between ways of selecting the model). 

When the model space to be considered is much larger than the data which is available, the standard approach is to modify the objective function in order to 'regularize' the model space and thereby prevent overfitting. The idea is that rather than treating all points in the model space as equally good, one can choose to rank them in some order (presumably in order of complexity), and then favor low complexity solutions over higher complexity solutions which are equally good. Or more flexibly, one can assign a cost to complexity and add it to the objective function. One way that this is done in the case of linear models is to assign a cost to model weights being large, corresponding to an inductive bias that the functions one is trying to learn should be smooth. A further constraint might be that the number of non-zero coefficients should be small. These are, respectively, $L_2$ and $L_1$ regularization. It turns out that $L_1$ regularization can reduce the amount of data needed from linear in the parameter count to being only logarithmic \citep{ng2004feature}.

Neural networks generally have far more parameters than there is data to train them. Furthermore, there are tight bounds which tie the parameter count to the VC dimension \citep{bartlett2017nearly}. Specifically, the VC dimension $d_{vc}$ is bounded by:

\begin{equation}
c_1 WL \log(W/L) \leq d_{vc} \leq c_2 W\bar{L} \log(2U)
\end{equation}

where $W$ is the number of parameters, $L$ is the network depth, $\bar{L}$ is the 'average' network depth weighed by parameter count, and $U$ is the number of nonlinear units and is proportional to $W$. Modern image classification neural networks have parameter counts in the hundreds of millions, but are generally trained on a standard dataset (ImageNet) which has only 1 million images. 

As such, it appears that neural networks operate in a regime which would require strong regularization in order to obtain any guarantees that the relations they learn correspond to actual systematic structures underlying the distributions of different types of natural images. To this end, $L_1$ and $L_2$ regularization along with a number of stochastic regularization methods such as Dropout \citep{srivastava2014dropout} were used extensively with respect to $5\times10^4$ image datasets such as MNIST and CIFAR to obtain state of the art performance. However, these methods have become less and less common for large-scale data even on the level of ImageNet, much less the $3\times10^8$ or $3.5\times10^9$ image datasets. If we think about this from the perspective of open-endedness requiring a suppression of characteristic scales, this makes sense because any regularization term added to the objective function which penalizes complexity would establish some characteristic scale beyond which the reward for putting more information into the network would be less than the corresponding penalty being assigned. 

Curiously, even without explicit regularization, neural networks seem to overfit far less than their VC dimension would suggest that they could. Furthermore, there is evidence that increasing the parameter count can actually increase generalization performance in practice \citep{neyshabur2017implicit}. While this suggests that the VC dimension generalization bound is simply not a tight bound on the learning process used to train neural networks, it has been observed that when the labels are randomized (destroying any systematic relationship between the inputs and targets of prediction), networks can in fact still learn to memorize the mapping between individual images and those arbitrary labels \citep{zhang2016understanding}.

It appears that what is going on is a form of regularization that is inherent to the training process itself \citep{neyshabur2017implicit}. Unlike regularization applied by modifying the objective function, this implicit regularization does not necessarily sacrifice the ability to represent or even find arbitrarily high-complexity states (as seen in the observation that neural networks can ultimately just memorize their training data if there are no other patterns to find). But rather, it must take the form of a preference in the order for which solutions are explored. Thus, if a solution with zero training error and good generalization performance exists, it has an increased chance of being favored over a solution with zero training error and bad generalization performance, so long as the inductive bias associated with the training process has some structural commonality with the type of problems that exist out in the world. 

This gives us a self-tuning property that may be advantageous in looking for systems whose complexity increases due to internal process rather than by being driven by an external schedule. That is to say, for an organism embodied within a system and implementing such a learning mechanism, both the depth of model being searched and the amount of data available to condition the model would scale together in time. As long as more data (e.g. interactions with the world) are constantly being added, then the generalizable complexity bound and also the depth into the parameter space searched by stochastic gradient descent can both scale in parallel.

\subsection{Generative Complexity}

There is a continuing debate in ALife as to whether the complexity of biological systems arises primarily from a complexifying process inherent in evolution itself, or if it is due to nascent complexity and richness which exists within the environment life finds itself in (or, more abstractly, in the laws of chemistry and physics) \citep{rasmussen2001ansatz}. That is to say, one hypothesis for why artificial life systems exhibit bounded complexity is that hand-constructed artificial systems tends to be much cleaner than what you'd find in a random spot on Earth --- be it due to fluctuations such as seasons, rare events, transport from surrounding regions, heterogeneity of composition, etc. Similarly, the sorts of rules constructed in toy models to try out ideas tend to have less inherent variation than one would observe in looking at things such as real chemical reaction networks. On the other hand, there is an argument that even if such things have a great degree of richness and might help drive the complexity of life, that richness had to come from somewhere fundamental --- ultimately chemistry derives from quantum mechanics, which does not possess a separate fundamental constant for each chemical species or chemical reaction in the network, and similarly the richness of a real environment derives from fundamentally simpler underlying phenomena such as chaos. The previous examples of images, sounds, and language are all cases with external datasets. We have shown that neural networks can adapt to complexity provided to them over a range of scales, sometimes in a coevolutionary manner, but we have not yet looked at whether neural networks are suitable vehicles for generating complexity increases on their own due to their internal dynamics.

However, results involving \emph{de novo} self-generated complexity are starting to appear in the machine learning literature. In particular, recent approaches involving reinforcement learning make heavy use of the idea that one can bootstrap sophisticated strategies by having a network play games against copies of itself. In cases where the rules of the game are known exactly and the game state is fully visible, the current leading method is expert iteration \citep{anthony2017thinking}. The core idea revolves around the construction of an operator which takes a probabilistic gameplay policy as input and returns a policy which is at least as good as that policy (but is capable of being better). Because the game rules are known (even if stochastic) and all information is available to all players, it is possible to use strong theoretical guarantees from Monte-Carlo Tree Search \citep{browne2012survey} about bounded regret in order  to produce such an operator. Given such a policy improvement operator, the training method is simply to train each agent to imitate its better self. Over a comparatively short training period of months (versus the thousands of years that humanity has studied the game), this method has produced agents which have beaten top human players at Go \citep{silver2016mastering, silver2017mastering} and similarly demonstrates better performance than top Shogi and Chess engines. So at least within the scope of strategies implied by the (fixed) rules of such games, networks coupled with Monte-Carlo Tree Search are capable of discovering more complexity than they are directly provided. 

In partially observable games, as well as games where the rules are not made available to the network, work is ongoing to determine the degree to which reinforcement or other approaches can discover similar levels of nested complexity. Various strategies have been discovered in the context of simulated physical competitive tasks such as wrestling \citep{bansal2017emergent}. Projects to map reinforcement learning into real-time strategy game environments such as Starcraft and DOTA 2 are ongoing (with a recent exhibition match by a set of agents produced by OpenAI appearing to approach the level of professional play, though the details are as of yet unpublished at the time of writing). So even without the underlying conditions for using expert iteration, it appears that some degree of complexity can be discovered.

In parallel to the contrast between internally and externally prompted complexification, there are a number of approaches to formulate 'intrinsic' (as opposed to externally imposed) motivation functions. These comprise concepts such as minimization of surprise \citep{friston2015active}, maximization of empowerment \citep{klyubin2005empowerment}, and curiosity \citep{schmidhuber1991possibility, bellemare2016unifying}. It is interesting to observe that, much like how many artificial evolutionary systems seem as though they could go open-ended but instead saturate around some fixed-complexity set of solutions, a recurring challenge behind formulating intrinsic motivations is the occurrence of a so-called 'dark room problem' \citep{little2013maximal}, in which there is some trivial way in which an agent following that motivation can globally maximize it without significant effort or engagement with the actual dynamics of the environment. An example of this is that an agent which attempts to minimize it's surprisal could conceivably learn about the world and make an advanced predictive model, but instead it would be simpler for it to find ways to turn off it's senses and thereby avoid all sources of potential surprise.

Dark room problems and their corresponding solutions seem to share some commonality with the issues surrounding convergence of evolutionary processes to a finite complexity fixed point, versus (co-evolutionary) dynamics which can in some cases lead to open-ended arms races. It has been observed that intrinsic motivations can be grouped into homeostatic and heterostatic cases --- where homeostatic motivations admit a fixed point stationary behavior that globally optimizes the motivation function, while heterostatic ones have no stable fixed points \citep{oudeyer2009intrinsic}. In homeostatic cases, the solution to dark room problems is to impose constraints which are non-trivial to satisfy, such that any complex structure which emerges takes its shape from the interaction between the constraint and the motivation function. In the context of surprise minimization, this could take the form of imposing a prior belief that a given outcome will be achieved. On the other hand, heterostatic intrinsic motivations make use of internal tensions to prevent any single policy or strategy from being stable. A curiosity-driven agent, for example, might simultaneously be trying to predict something accurately and be trying to take actions which make its own predictive models fail \citep{pathak2017curiosity, burda2018exploration, yu2018boredom}. In comparison, evolutionary dynamics where fitness is a fixed function of the individual genotype of each organism on its own converge to a stationary distribution around optima (essentially homeostatic), whereas co-evolutionary dynamics are capable of expressing non-stationary dynamical outcomes such as limit cycles, chaotic behavior, or travelling wave type solutions (heterostatic) and as a result are more able to avoid getting stuck in ways that inhibit open-endedness. 

\section{Shifts in individuality}

The things we have mentioned still constitute examples of fixed games, so it is likely that there is some upper bound on the degree of useful complexity which can ever be discovered even by the best possible population of Go players or Starcraft players. Furthermore, within the context of a fixed framework such as Go or Starcraft, there is no way for the type of emergent complexity to extend to modes of interaction far beyond the game itself. A reinforcement learning agent trained on Starcraft does not possess the capacity to, for example, decide to take up knitting.

On the other hand, real biological systems often breach their bounds in surprising ways, seemingly changing the rules of the game. These events happen at the small scale, such as the creation of new niches or cultivation of aspects of the environment, but also at the scale of redefining the unit of individuality in a succession of major transitions throughout the history of life on Earth \citep{smith1997major}. For example, we can contrast the sort of results we would expect if we began from a hard constraint of modelling life as a well-mixed chemical reaction network, versus what actually ended up happening. In a well-mixed chemical system, environmental conditions such as temperature and pH will dominate what chemistry is possible, and so while one might have autocatalytic sets or hypercycles \citep{eigen2012hypercycle}, they would be at the mercy of that external driver. By making use of membranes, a physical process depending on the spatial arrangement of molecules (and so invisible to a description of the world as a well-mixed one-pot reactor), organisms can achieve homeostasis and partially decouple the internal chemical environment from the external one. When reactions being always on or always off would put strong limits on the stability of cycles (such as in the Eigen hypercycle case \citep{eigen2012hypercycle}), enzymes, gene expression, and other things can be invented in order to again change the assumptions and effective rules of the game. When limits of what chemistry can take place in a shared space arise, grouping, symbiosis, or parasitism between individuals can lead to colonies and eventually multicellularity, changing the fundamental unit of individuality in the system.

If ultimately we want an open-endedness of that kind rather than simply open-endedness driven by the complexity of a fixed world, we need to consider both ways in which the rules of the game can themselves be altered, and ways in which agents can modify their effective boundaries and the way in which their identities are encoded and expressed. This is quite a lot to ask for in terms of any artificial system, but one approach has been to try to resolve the underlying operations into physically embodied processes. That is to say, in something like Tierra, replication is not assumed as a rule of the system, but rather must be accomplished via emergence from lower-level pieces --- meaning that the nature of replication itself could change.

In evolution, there is a tension between scales in that each self-replicating component on its own might evolve in such a way as to increase its local fitness at the expense of the fitness of the macro-structure of which it is a part. Mechanisms and forms of biological order exist which suppress this tendency --- the use of a bottleneck germ cell to mediate replication of the whole, programmed cell death, etc. In essence, there must be some way in which information about the structure of the entire organism at the higher scale propagates and influences the behavior of each individual component (through having shared genes, through top-down regulation, through sharing common bottleneck points in time, etc).

When it comes to neural networks, we can think of the lower-level pieces as being individual parameters or mathematical operations. The computation executed by a network is composed by those contributions working together to construct some overall function, and the 'individual' could be considered either to be each of those parameters, or the network as a whole. In that sense, there is already some degree to which the behavior of the network as a cognitive system is dependent on it crossing a boundary between the behavior of each component and the interface between the whole and whatever task it is trained on. In backpropagation-based gradient descent, this is obtained by analytically computing the derivative of the global behavior (as defined through the lens of the objective function) with respect to each parameter --- in effect, providing each parameter a summary statistic representing what the downstream degrees of freedom 'need' in order to make the overall behavior change in a certain targetted way. Although this summary statistic contains global information about the structure of the network as a whole, it can be computed through a series of entirely local operations (which amount to repeated applications of the chain rule). As a result, backpropagation fills the role of evolutionary mechanisms such as group selection, in that it provides the necessary coupling between scales and causes the individual parameters to behave collectively so as to produce a coherent overall response to the external task. 

In practice, neural architectures are hand-specified and fixed with respect to a particular problem and course of training. This means that, while the framework of a backpropagation-based neural network can include many different forms of macro-scale 'individual', most work concerns itself with particular forms that can take rather than transitions between them. There are some exceptions however: methods such as NEAT \citep{stanley2002evolving} use evolutionary methods to allow network architectures to adapt in response to a problem, neural architecture search \citep{zoph2016neural} uses reinforcement learning to learn a probabilistic policy for constructing new architectures, and adaptive neural trees \citep{tanno2018adaptive} recursively and dynamically generates a neural network architecture on the fly as it learns.

We can also consider cases in which the logical structure of a computation changes dynamically even if the architecture is fixed. Some examples of this are neural Turing machines \citep{graves2014neural, graves2016hybrid} and memory-augmented neural networks \citep{santoro2016meta}. In these cases, the network's input is arbitrarily extensible, either in the form of an input sequence or in the form of a preloaded external memory whose size can vary. An attention mechanism is used in order to force the network to, internally, decide upon an execution path and access pattern over the data available to it. This means that, even using the same fixed architecture for the network, one could have computations of varying length and extent over the data.

The above methods might be thought of as the network changing and diversifying its' own interior structure, essentially redefining the sense of the individual 'inwards'. However, the actual motivation and task is still held constant and fixed --- all that happens is that the system organizes in different ways to respond to that pressure. On the other hand, if we want a situation in which the rules of the game change fundamentally, the objective function or pressures must also be part of those dynamics. We can look at something like the GAN architecture as having an aspect of that --- by having one network provide a supervision signal to another, the generator's objective becomes a dynamic feature of the system as a whole rather than a fixed external thing. Work has been done on training populations of interacting networks in cooperation games, such as \citep{foerster2016learning} which uses differentiable communication channels to backpropagate supervision signals between a pair of interacting agents playing a coordination game, and \citep{mordatch2017emergence} which extends that to the case of agents learning to develop a shared, compositional language. From the point of view of the backpropagation pass, these multi-agent setups just correspond to something that is, instantaneously, a single network that just happens to have a dynamically varying architecture. 

One could go even further and allow the networks to determine on their own how to associate with one-another. While different communication topologies are on the face of it a binary choice, the same sort of continuous extension of that which is provided by attention mechanisms (such as \citep{vaswani2017attention}) can in principle render the choice of which sub-network to receive from into a differentiable (and therefore backpropagation-compatible) function. The idea in this case would be that each agent and associated sub-network represents the others around it as a vector by observing their behavior (as in \citep{rabinowitz2018machine}), and then uses those vectors in order to choose which to associate with or pull information from.  This would enable a neural-network-based system that could simultaneously represent dissociated individuals, collectives, and the cognitive process by which a transition between them occurs. In such a case, the supervision signal might be externally defined at the level of individual subnetworks, but when those signals are incompatible with each-other (as in a GAN) the system as a whole can behave in a way which emerges from the interaction of those individual motivations with the learned associations between sub-groups within the population --- in essence, giving rise to new emergent, population-level motivations.

The above approaches allow the internal and external arrangements of a neural-network-based system to become dynamic, but the learning algorithm itself is still fixed. However, there are a number of ways in which this constraint may be relaxed. The concept of 'neuromodulation' captures the ideas that the parameters of adaptation or learning may, themselves, adapt on a slower timescale. This has been used to learn control schemes for learning rate on reinforcement learning tasks \citep{cussat2015genetically}. Beyond just tuning the parameters of a fixed learning algorithm, the entirety of the learning algorithm may be made subject to adaptation. These approaches fall under the general heading of 'learning to learn' \citep{hochreiter2001learning, bosc2016learning, andrychowicz2016learning}, where the functional form of the weight updates of a network is discovered via some other optimization algorithm --- this can be a genetic algorithm, reinforcement learning, or even backpropagation through the learning algorithm itself. A recent example of this idea uses backpropagation through the gradient descent procedure to train networks to be able to quickly learn when presented with new tasks, enabling one-shot learning in classification and robotics control task domains \citep{finn2017model, finn2017one}. There is also recent work looking at the possibility of training networks to train each-other (such as by learning curricula \citep{misra2017learning} or data augmentation strategies \citep{cubuk2018autoaugment}). If differentiable communication can be thought of as allowing networks to alter the boundary of individuality over space, these meta-learning techniques in some sense allow the alteration of the boundary of individuality over (learning) timescales.

Ultimately, while we cannot yet answer the question of what it would take to achieve an open-ended succession of multiple major transitions as biological evolution appears to have done, there are a number of techniques which could be used within the framework of machine learning and neural networks to probe the possibility of emergent shifts in the definition of the individual. The inherent recursivity of a backpropagation pass seems to lend itself towards being able to fluidly vary the scale on which motivations are compartmentalized while retaining the same basic underlying information. As such, we hope to encourage researchers in artificial life to explore both ways in which this may be taken advantage of, and ways in which these structures can be understood in light of insights garnered from working with evolutionary analogues.

\section{Conclusions}

Understanding the open-ended complexity of the natural world is one of the greatest long-standing problems evolutionary biology, and emulating such complexity is one of the greatest open challenges in Artificial Life. While most would agree that a complete understanding is yet to come, substantial progress has been made through evolutionary simulations. At the same time, the field of machine learning is in an era of rapid progress, with neural networks reaching levels of functional complexity that were undreamed of only a decade ago. We have argued that several emerging ideas and techniques from that field can be brought to bear on questions about open-ended evolution and, \emph{vice versa}, that ideas about coevolutionary complexity are becoming increasingly relevant within machine learning itself.

We have reviewed a number of topics relating to the emerging crossover between these two fields, focusing in particular on the issues of diversity and scaling. We believe that this convergence of ideas will provide substantial new insights to both fields in the years to come.

\section*{Acknowledgements}

The authors would like to thank Martin Biehl for helpful discussions. Nathaniel Virgo is partially supported by by JSPS KAKENHI Grant Number JP17K00399. We are grateful to the Earth-Life Science Institute (ELSI) for funding a visit by Alexandra Penn.

\bibliographystyle{apalike}
\bibliography{oee}

\end{document}